\documentclass[aps,prb,twocolumn,superscriptaddress,showpacs,floatfix]{revtex4}
\usepackage{hyperref}
\usepackage{amsmath,amssymb,amsfonts}
\usepackage{graphicx}
\usepackage[percent]{overpic}
\DeclareGraphicsExtensions{.jpg,.pdf,.png,.eps}
\newcommand{\znfe} {YbFe$_{2}$Zn$_{20}$ }
\newcommand{\znfef} {YbFe$_{2}$Zn$_{20}$}
\newcommand{\Tk} {$T_{\mathrm{K}}$ }
\newcommand{\Tkf} {$T_{\mathrm{K}}$}
\newcommand{\Tfl} {$T_{\mathrm{FL}}$ }
\newcommand{\Tflf} {$T_{\mathrm{FL}}$}
\newcommand{\Tsq} {$T^{2}$ }
\newcommand{\Tsqf} {$T^{2}$}
\newcommand{\Tmax} {$T_{\mathrm{max}}$ }
\newcommand{\Tmaxf} {$T_{\mathrm{max}}$}

\newcommand{\TCf} {$T_{\mathrm{C}}$}
\newcommand{\Pc} {$P_{\mathrm{c}}$ }

\newcommand{\Pen} {1~:~1 n-pentane~:~iso-pentane }

\newcommand{\minoilf} {6~:~4 n-pentane~:~mineral~oil}
\newcommand{\FC} {1~:~1 FC70~:~FC770 }
\newcommand{\FCf} {1~:~1 FC70~:~FC770}

\begin{document}
\title{Search for Pressure Induced Quantum Criticality in YbFe$_{2}$Zn$_{20}$}
\author{S. K. Kim}
\affiliation{Ames Laboratory, Iowa State University, Ames, Iowa 50011, USA}
\affiliation{Department of Physics and Astronomy, Iowa State University, Ames, Iowa 50011, USA}
\author{M. S. Torikachvili}
\affiliation{Ames Laboratory, Iowa State University, Ames, Iowa 50011, USA}
\affiliation{Department of Physics, San Diego State University, San Diego, California 92182, USA}
\author{S. L. Bud'ko}
\affiliation{Ames Laboratory, Iowa State University, Ames, Iowa 50011, USA}
\affiliation{Department of Physics and Astronomy, Iowa State University, Ames, Iowa 50011, USA}
\author{P. C. Canfield}
\affiliation{Ames Laboratory, Iowa State University, Ames, Iowa 50011, USA}
\affiliation{Department of Physics and Astronomy, Iowa State University, Ames, Iowa 50011, USA}
\date{\today}

\begin{abstract}
Electrical transport measurements of the heavy fermion compound YbFe$_2$Zn$_{20}$ were carried out under pressures up to 8.23~GPa and down to temperatures of nearly 0.3~K.  The pressure dependence of the low temperature Fermi-liquid state was assessed by fitting $\rho\mathrm{(}T\mathrm{)}=\rho_0+AT^n$ with $n$~=~2 for $T<T_{\mathrm{FL}}$.  Power law analysis of the low temperature resistivities indicates $n$~=~2 over a broad temperature range for $P\lesssim 5$~GPa.  However, at higher pressures, the quadratic temperature dependence is only seen at the very lowest temperatures, and instead shows a wider range of $n<2$ power law behavior in the low temperature resistivities.  As pressure was increased, $T_{\mathrm{FL}}$ diminished from $\sim 11$~K at ambient pressure to $\sim 0.6$~K at 8.23~GPa.  Over the same pressure range, the $A$ parameter increased dramatically with a functional form of $A\propto \mathrm{(}P-P_{\mathrm{c}}\mathrm{)}^{-2}$ with $P_\mathrm{c}\simeq 9.8$~GPa being the critical pressure for a possible quantum critical point.  
\end{abstract}

\pacs{62.50.-p,72.15.Eb,72.15.Qm,75.20.Hr,75.30.Kz}
\maketitle 

\section{Introduction}
Strongly correlated electron systems manifest a rich variety of electronic and magnetic properties that have fascinated scientists for decades; Mott insulators,\cite{Mott-MIT90} exotic superconductors,\cite{Bennemann-08,Norman-SM11} and heavy fermions\cite{Stewart-RMP84,Stewart-RMP01,Stewart-RMP06} continue to be topics of intense study.  Heavy fermions are a subset of materials known as Kondo lattices.  In a Kondo lattice system, near the characteristic Kondo temperature, \Tkf , the moment bearing ions in the lattice each act as a Kondo impurity, with which the conduction electrons hybridize and create a screening cloud of dynamically polarized electrons.  At lower temperatures, the conduction electrons become coherent and manifest Fermi-liquid (FL) behavior.  This leads to a myriad of novel features in measurements of the Kondo lattice's thermodynamic and transport properties at and below \Tkf : the susceptibility, which has a local moment-like, paramagnetic behavior at high temperatures, shows a loss of local moment behavior below \Tkf ; in measurements of specific heat, a large Sommerfeld coefficient for $T<T_{\mathrm{K}}$ points to a large effective electron mass; resistivity measurements for $T\ll T_{\mathrm{K}}$ are Fermi-liquid-like and are consistent with a large electronic density of states at the Fermi energy. 

The YbTM$_{2}$Zn$_{20}$ (TM~=~Fe, Co, Ru, Rh, Os, and Ir) compounds form a family of Yb-based heavy fermions where subtle changes in the hybridization of the Yb local moment with the conduction electrons can be achieved by changing the transition metal (TM) element.  These materials show classical heavy fermion behavior in their physical properties.\cite{Torikachvili-PNAS07,Canfield-PB08,Jia-PRB09,Jia-Thesis08,Mun-PRB12}  Of this series, YbCo$_2$Zn$_{20}$ seems to be an outlier with $T_K$ much lower than those of the other five members (\Tk =~33, 30, 16, 20, and 21~K for TM~=~Fe, Ru, Rh, Os, and Ir, respectively, and \Tk =~1.5~K for TM~=~Co).\cite{Torikachvili-PNAS07}  Specific heat measurements yielded large Sommerfeld constants ($\gamma$~=~520, 580, 740, 580, 540~mJ/mol~K$^{2}$ for TM~=~Fe, Ru, Rh, Os, and Ir, respectively, and an exceptionally large value of 7200~mJ/mol~K$^{2}$ for YbCo$_2$Zn$_{20}$).\cite{Torikachvili-PNAS07}  Magnetic susceptibility measurements show Curie-Weiss behavior at high temperatures and a broad maximum at low temperatures, signalling a Kondo screened local moment.  When the magnetic component of the resistivity is isolated by subtracting from it the resistivity of the non-magnetic counterpart LuTM$_2$Zn$_{20}$, the resistivities of all members of the YbTM$_{2}$Zn$_{20}$ series show a local maximum near \Tkf .  Below \Tkf , the coherent scattering of electrons decreases the resistivity until at very low temperatures, a wide temperature range of $T^2$ behavior signals the onset of Fermi-liquid behavior.\cite{Coleman-07}

Although no indication of low temperature, local moment order was found for any of the YbTM$_2$Zn$_{20}$ compounds,\cite{Torikachvili-PNAS07,Canfield-PB08,Jia-PRB09,Jia-Thesis08,Mun-PRB12} by combining the idea of the basic Doniach model\cite{Doniach-PB77} with simple steric arguments, it is anticipated that the application of pressure will stabilize local-moment-like states in Yb-based Kondo lattic systems.  Indeed under pressure, indications of a magnetic instability were seen in YbCo$_{2}$Zn$_{20}$.  When a modest pressure of $\sim 1$~GPa was applied, the resistivity measurement showed an anomaly, possibly magnetic in origin, at $\sim 0.15$~K.\cite{Saiga-JPSJ08,Matsubayashi-JPCS10}  Given the drastic difference in Kondo temperatures between YbCo$_{2}$Zn$_{20}$ and \znfe a significantly larger amount of pressure is expected to be needed to drive the \znfe system to a magnetically ordered state at low temperature.  For YbCo$_{2}$Zn$_{20}$, at the critical pressure needed to induce an ordered state, the temperature of the local maximum of the magnetic component of the resistivity (\Tmaxf ), was at its minimum value.  A similar minimum was also seen in pressure measurements on YbRh$_{2}$Zn$_{20}$ and YbIr$_{2}$Zn$_{20}$ at $\sim 3$~GPa,\cite{Matsubayashi-JPCS10, Matsubayashi-JPCS09} however, no features signalling magnetic ordering were seen down to 2~K for these compounds.  This is not surprising considering the magnetic ordering temperature for YbCo$_{2}$Zn$_{20}$ was, at most, $\sim 0.5$~K at 2.37~GPa.  If indeed both YbRh$_{2}$Zn$_{20}$ and YbIr$_{2}$Zn$_{20}$ with \Tk =~16 and 21~K, respectively, have critical pressures near 3~GPa, then it is expected that for \znfef , which has a higher Kondo temperature of \Tk =~33~K,\cite{Torikachvili-PNAS07} the critical pressure for a pressure induced QCP will be $P_{\mathrm{c}}>3$~GPa.  On the other hand, \znfe may well manifest a higher magnetic ordering temperature than YbCo$_{2}$Zn$_{20}$.  A comparison of the relative ordering temperatures of GdCo$_{2}$Zn$_{20}$ and GdFe$_{2}$Zn$_{20}$\cite{Jia-NP07,Jia-PRB09} combined with de Gennes scaling suggests that if an RKKY interaction induces a magnetically ordered state for the Yb system, a magnetic transition temperature, as high as $\sim 1$~K may be found for \znfe under pressure.  Moreover, other members of the family of RFe$_{2}$Zn$_{20}$, namely R~=~Gd-Tm manifest ferromagnetic ordering which suggests that the ordered state in \znfef , if it can be stabilized, may be more complex than simple AFM order.

%-------------------------------------------------------------------------------
%-------------------------------------------------------------------------------

\section{Experimental methods}
Starting with a high temperature melt of stoichiometry: Yb:Fe:Zn~=~2:4:94, single crystals of \znfe were grown out of excess Zn.  A detailed description of this process is described elsewhere.\cite{Torikachvili-PNAS07,Jia-NP07,Canfield-PB08,Jia-PRB09,Canfield-Euroschool10}

For resistivity measurements, the single crystals were polished down to appropriate dimensions for the two types of pressure cells used: $1.6\times 0.5\times 0.4$~mm$^3$ for the piston cylinder cell ($P\lesssim$~2~GPa)\cite{Torikachvili-PRB08,Torikachvili-PRB09} and typically $700\times 150\times 30~\mu \mathrm{m}^3$ for the modified Bridgman anvil cells (mBAC) (2.85~$\leq P \leq 8.23$~GPa).\cite{Colombier-RSI07}  The small sizes of the samples used in these pressure cells lead to relatively large geometric errors in the resistivity values.  Therefore, all resistivity measurements were normalized to that of a single measurement on a large sample which was used in a previous study on the transport properties of Yb-based heavy fermions.\cite{Jia-Thesis08,Jia-PRB09}  The normalization was done by multiplicatively matching the resistivity at ambient pressure and at 298~K, then applying the same multiplicative value to the rest of the resistivity measurements under pressure.  Given that these materials have cubic unit cells,\cite{Torikachvili-PNAS07,Jia-NP07,Canfield-PB08,Jia-Thesis08,Jia-PRB09} this normalization procedure is considered to be quite reliable.

The piston cylinder cell was used for pressure measurements up to about 2~GPa.\cite{Torikachvili-PRB08,Torikachvili-PRB09}  This cell was designed to be used in a Quantum Design Physical Properties Measurement System~(PPMS) with the dc resistivity option.  The cell was fitted with a NiCrAl inner core and the pressure medium was a mixture of \minoilf .  At room temperature, the pressure inside the cell was monitored by using the resistance of a Manganin wire and at low temperatures, the superconducting transition temperature of a lead sample was used.\cite{Eiling-JPF81}

A modified Bridgman anvil cell was used for resistivity measurements at pressures up to about 8.2~GPa.  This cell has been designed to work inside a PPMS and has been modified to use a liquid pressure medium.\cite{Colombier-RSI07}  For this study, two pressure media were used with the mBAC:  a mixture of \Pen and a mixture of 1~:~1 Fluorinert 70~:~Fluorinert 770 (\FCf ).\cite{Piermarini-JAP73,Klotz-JPDAP09,Varga-RSI03,Sidorov-JPCM05} Despite the the lower pressure of solidification at room temperature of the latter liquid medium, its lower compressibility allowed higher pressures to be more readily achieved.  As in the piston cylinder cell, the superconducting transition temperature of a lead sample was used as a pressure gauge within the mBAC sample space.

At ambient pressure, \znfe samples had residual resistivity ratios that varied from 18 to 38.  The low-temperature resistivity follows $\rho\mathrm{(}T\mathrm{)}=\rho_{0}+AT^2$ with the value of $A$ having an average of $7.0\pm 0.3\times 10^{-10}~\Omega$~cm~K$^2$ and the crossover temperature, \Tflf , below which this form fits the data, has an average value of $11.4\pm 0.8$~K.  With two different pressure cells and three different liquid media, there are clear, albeit small, differences between the various pressure dependent data sets, that may (in part) be attributed to cell to cell and/or sample to sample differences but these are small compared to the much larger changes due to pressure (see Fig.~\ref{ZnFL_PD} below and associated discussion).

For low temperature measurements, down to nearly 0.3~K, a CRYO Industries of America $^{3}$He system was used to measure samples under pressure with the mBAC.  

For most pressure cells, one cycle of cooling and warming causes a pressure cell to thermally contract and expand.  One subtlety of pressure measurements with the mBAC is that this thermal cycling can induce a modest pressure increase within the pressure cell.  This is the origin of the slight pressure differences between subsequent measurements with the PPMS (used for measurements down to 2~K) and the $^{3}$He system (used for measurements down to nearly 0.3~K).  An example of this effect is seen below in Fig.~\ref{ZnPaT2}.  

%-------------------------------------------------------------------------------
%-------------------------------------------------------------------------------

\section{Results}

\begin{figure}[!htbp]
\centering
\includegraphics[width=80mm]{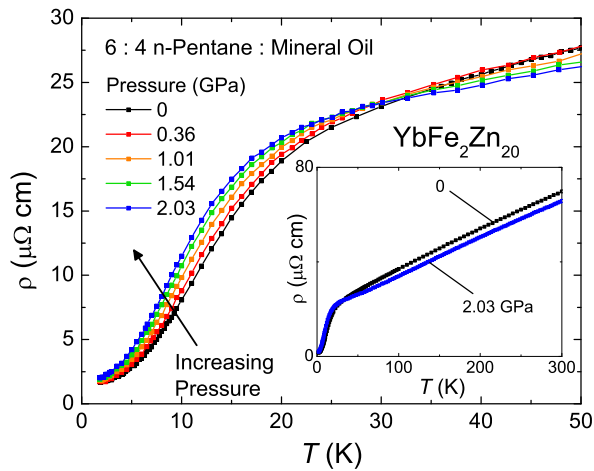}
\caption{(Color Online)~Temperature dependent resistivity measurement of \znfe under pressure using a piston cylinder cell are shown up to 50~K for pressures up to 2.03~GPa.  Inset:~Full temperature range of the resistivity at ambient pressure and 2.03~GPa.  The high temperature region shows typical metallic behavior with a near-linear dependence on temperature.}
\label{ZnM}
\end{figure}

Temperature dependent resistivity measurements of \znfef , at ambient pressure and under pressures up to 2.03~GPa, are shown in Fig.~\ref{ZnM}.  These measurements were taken using a piston cylinder cell.  As temperature is decreased from 300~K (inset of Fig.~\ref{ZnM}) the resistivity decreases in a near-linear manner until, near 30~K, a broad shoulder-like drop appears.  Below this shoulder, the resistivity decreases rapidly with temperature and at the lowest temperatures manifests \Tsq behavior.  As the applied pressure is increased, the shoulder shifts to lower temperatures and the resistivity data for temperatures just above the shoulder flattens.

\begin{figure}[!htbp]
\centering
\includegraphics[width=80mm]{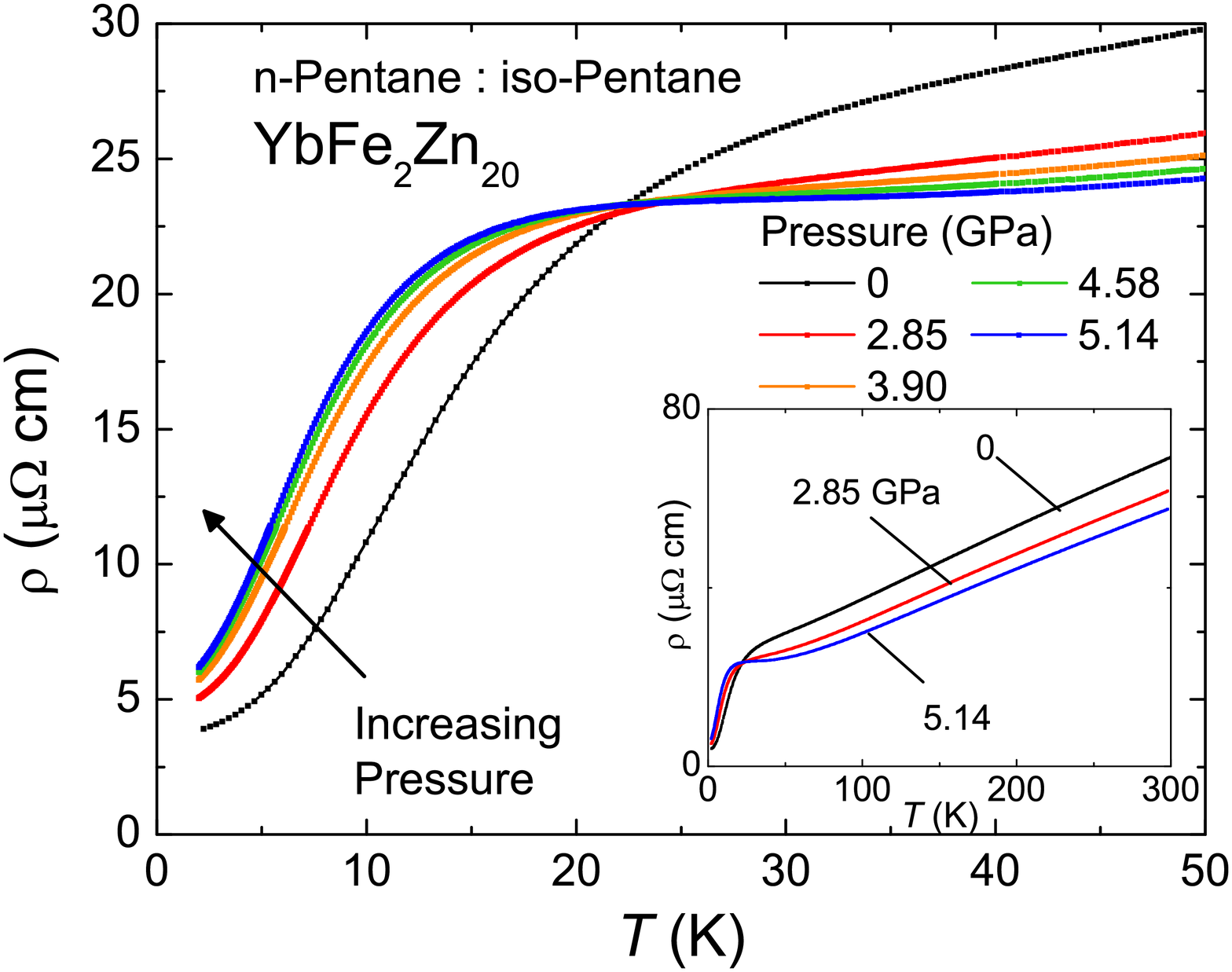}
\caption{(Color Online) Low temperature resistivity measurements under pressure using a modified Bridgman anvil cell with \Pen as the liquid pressure medium, reaching a maximum pressure of 5.14~GPa.  Inset: Resistivity curves for the full temperature range up to 300~K for pressures at 0, 2.85, and 5.14~GPa.}
\label{ZnPa}
\end{figure}

For measurements at higher pressures, the mBAC was used with both the \Pen mixture and the \FC mixture.  Figure~\ref{ZnPa} shows resistivity measurements with an mBAC for pressures up to 5.14~GPa using \Pen as the liquid pressure medium.  As pressure is applied, a local maximum in the resistivity above the shoulder grows as the shoulder itself shifts to lower temperatures.  

\begin{figure}[!htbp]
\centering
\includegraphics[width=80mm]{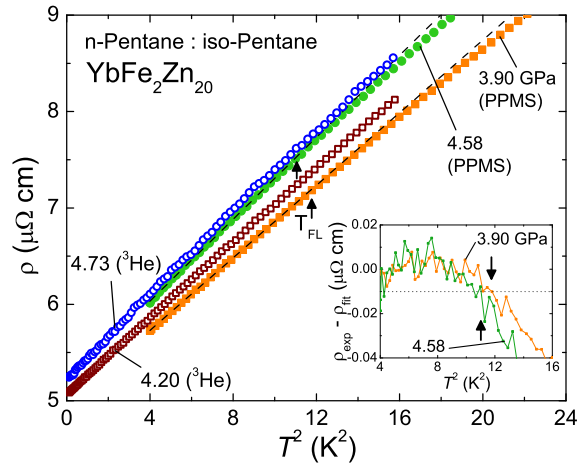}
\caption{(Color Online) \Tsq dependence of resistivity at low temperatures for measurements with the PPMS at 3.90 and 4.58~GPa shown as the closed squares and circles, respectively, as well as measurements with the $^3$He cryostat at 4.20 and 4.73~GPa denoted by the open squares and circles, respectively.  The black dashed lines are linear fits to the low temperature data taken with the PPMS.  The arrows indicate the temperature limit of \Tsq behavior, \Tflf , from PPMS data.  Inset:~$\rho_\mathrm{exp}-\rho_\mathrm{fit}$ for both 3.90 and 4.58~GPa versus \Tsqf.  The black dotted line indicates a difference of $-0.01~\mu\Omega$~cm used to determine \Tflf .}
\label{ZnPaT2}
\end{figure}

The $\rho\mathrm{(}T\mathrm{)}=\rho_0+AT^2$ behavior that persisted up to about 11~K at ambient pressure is diminished to a lower temperature range and the upward curvature becomes steeper as pressure is increased.  Measurements with the PPMS down to 2~K show quadratic behavior up to $\sim 3.3$~K at 4.58~GPa (Fig.~\ref{ZnPaT2}).  \Tfl was defined as the temperature at which the difference between experimental resistivity data, $\rho_\mathrm{exp}$, and linear fits to the $\rho_\mathrm{exp}$ versus \Tsq data, $\rho_\mathrm{fit}$, became greater than 0.01~$\mu\Omega$~cm.  For pressures where \Tsq behavior was still observable above 2~K, \Tfl was determined from measurements with the PPMS.  For \Tfl lower than 2~K, $^3$He data were used for the quadratic fits.

In order to explore the possibility of magnetic ordering at lower temperatures, measurements of resistivity down to nearly 0.3~K for 4.20 and 4.73~GPa were taken using a $^3$He cryostat, shown in Fig.~\ref{ZnPaT2}.  These measurements revealed no resistive anomalies down to almost 0.3~K and confirmed that \Tsq behavior persists down to our lowest temperatures.  
 
\begin{figure}[!htbp]
\centering
\includegraphics[width=80mm]{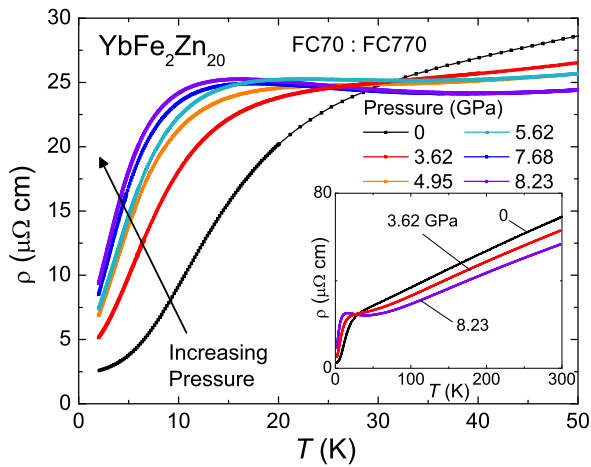}
\caption{(Color Online) Low temperature resistivity data for \znfe under pressures up to 8.23~GPa using \FC as the liquid pressure transmitting medium.  Selected curves are shown ($P$~=~0, 3.62, 4.95, 5.62, 7.68, and 8.23~GPa) from two separate sets of measurements.  Inset: Resistivity curves up to 300~K for $P$~=~0, 3.62, and 8.23~GPa.}  
\label{ZnF}
\end{figure}

In order to achieve higher pressures, the \FC liquid medium was used with the mBAC.  Figure~\ref{ZnF} shows selected resistivity curves for pressures up to 8.23~GPa from two different packings of the pressure cell (each using one half of the same crystal as a sample).  The resistivity in the high temperature region, between 30 and 300~K, continues to decrease with temperature (inset of Fig.~\ref{ZnF}) in a near-linear fashion.  The broad shoulder once again changes into a broad maximum with pressure and further increases up to 8.23~GPa reveal this maximum sharpening.  
 
\begin{figure}[htbp]
\centering
\includegraphics[width=80mm]{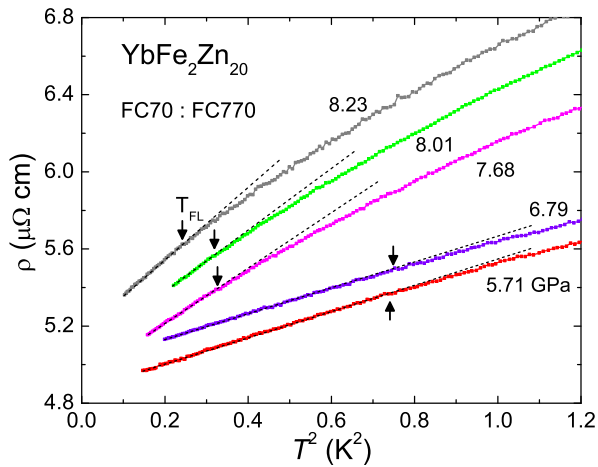}
\caption{(Color Online) \Tsq dependence of resistivity at pressures of 5.71, 6.79, 7.68, 8.01, and 8.23~GPa.  At maximum pressure, the \Tsq region has greatly diminished.  The black arrows indicate \Tfl values that were determined as shown in the inset of Fig.~\ref{ZnPaT2}.}
\label{ZnFT2}
\end{figure}

Measurements down to nearly 0.3~K were also conducted with \FC as the liquid medium in the mBAC (Fig.~\ref{ZnFT2}).  No features associated with magnetic ordering were seen in the data for pressures up to 8.23~GPa.  Above 5~GPa, the range of \Tsq behavior is suppressed to below 2~K.  This decrease in \Tfl continues with pressure and, at 8.23~GPa, the \Tsq behavior is found only up to $T\sim 0.6$~K (\Tsq$\sim 0.3$~K$^2$).  

%-------------------------------------------------------------------------------
%-------------------------------------------------------------------------------

\section{Discussion}
The application of pressure to Yb-based heavy fermion systems can shift the system from being dominated by the Kondo effect and manifesting low-temperature FL behavior, to being more local-moment like and even, in some cases, to manifesting long-range order.\cite{Alami-SSC98,Knebel-JPCM01,Trovarelli-PBCM02,Winkelmann-PRL98}  In resistivity measurements of a Kondo lattice system, the logarithmic rise of resistivity due to the Kondo effect is countered by the coherent scattering of the electrons at lower temperatures, creating a local maximum in the resistivity.  Since the Kondo temperature, \Tkf , sets an energy scale for a crossover from a local moment to a Kondo-screened moment state and is not correlated with any sharp features in physical measurements, it becomes necessary to isolate the magnetic contribution from the resistivity data in order to help determine, or set limits on, \Tkf .  It is expected that the position of this local maximum scales with \Tkf .\cite{Yoshimori-JMMM83,Coleman-JMMM87,Bauer-PRB93}

\begin{figure}[!htbp]
\centering
\includegraphics[width=80mm]{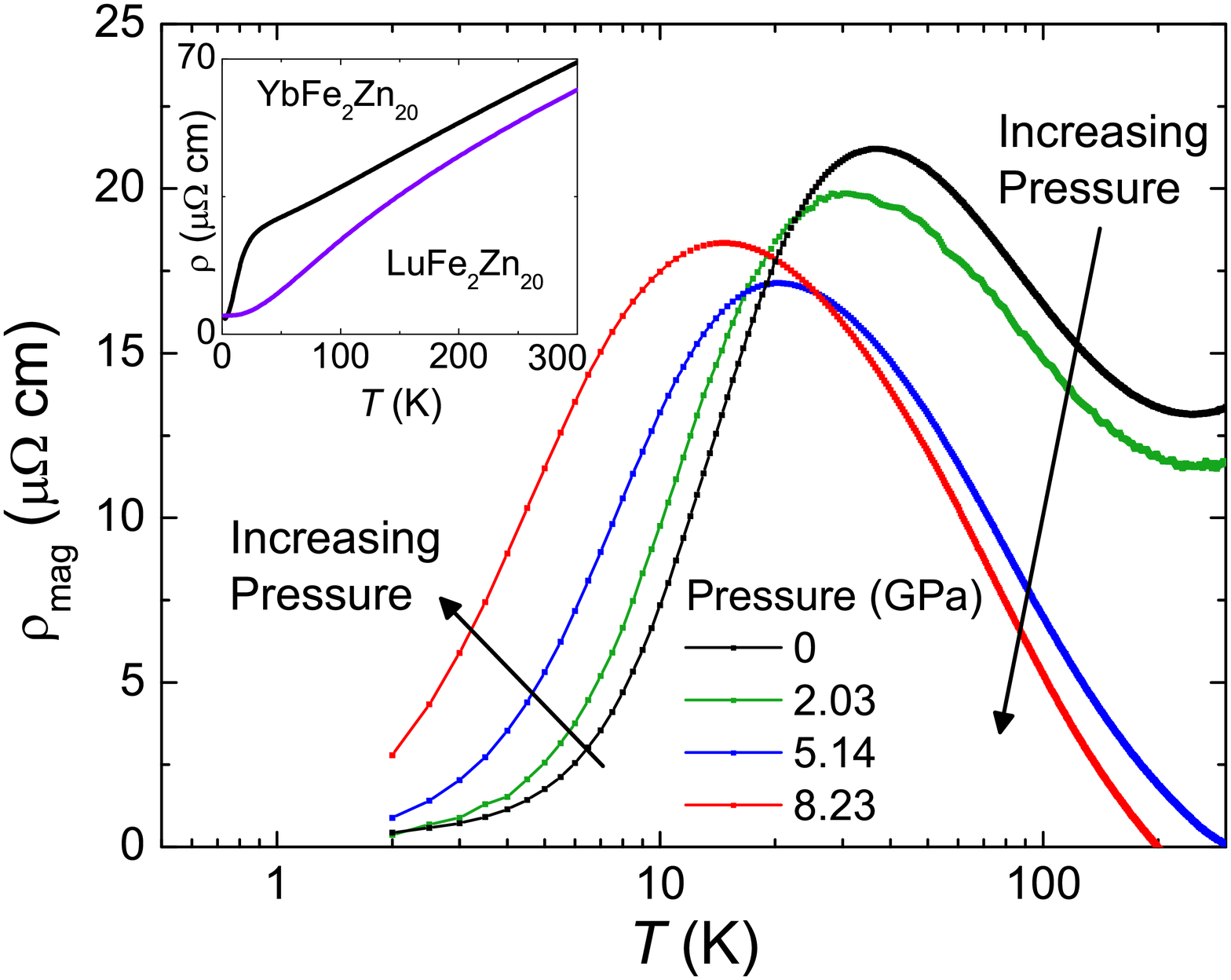}
\caption{(Color Online) The magnetic contribution, \mbox{$\rho_{mag}=(\rho-\rho_{0})_{\mathrm{Yb}}-(\rho-\rho_{0})_{\mathrm{Lu}}$}, to the total resistivity for various pressures are shown on a semi-log scale.  The inset shows the ambient pressure curves for \znfe and its non-magnetic analogue LuFe$_{2}$Zn$_{20}$.}  
\label{ZnRmag}
\end{figure}

The total resistivity of \znfe can be described as a combination of non-correlated, normal metal resistivity and a magnetic contribution to the resistivity.  The normal metal resistivity, can be approximated by the temperature-dependent resistivity of LuFe$_{2}$Zn$_{20}$, which, with a full 4$f$ shell, is a non-magnetic analogue to \znfef .  Both Yb and Lu curves are shown in the inset of Fig.~\ref{ZnRmag}.  By taking the difference, \mbox{$\rho_{mag}=(\rho-\rho_{0})_{\mathrm{Yb}}-(\rho-\rho_{0})_{\mathrm{Lu}}$}, the partially filled 4$f$ shell contribution to the resistivity can be effectively isolated.  This is shown for several measurements under pressure in Fig.~\ref{ZnRmag}.  For the subtraction, only the ambient pressure resistivity values of LuFe$_{2}$Zn$_{20}$ were used and is the likely cause of the negative $\rho_{mag}$ values at higher temperatures.

It is apparent that the low temperature maximum in the magnetic component of the resistivity decreases with pressure.  For Kondo lattice systems, the electrons participating in the spin-dependent screening in the region of the Kondo temperature are intimately related to the Fermi-liquid quasiparticles at low temperatures.  With previous results for YbCo$_{2}$Zn$_{20}$ having shown indications of a magnetic transition when \Tfl was suppressed,\cite{Saiga-JPSJ08,Matsubayashi-JPCS10} it is expected that a quantum critical point will also be reached when \Tfl in \znfe has been driven to zero.  

\begin{figure}[!htbp]
\centering
\includegraphics[width=80mm]{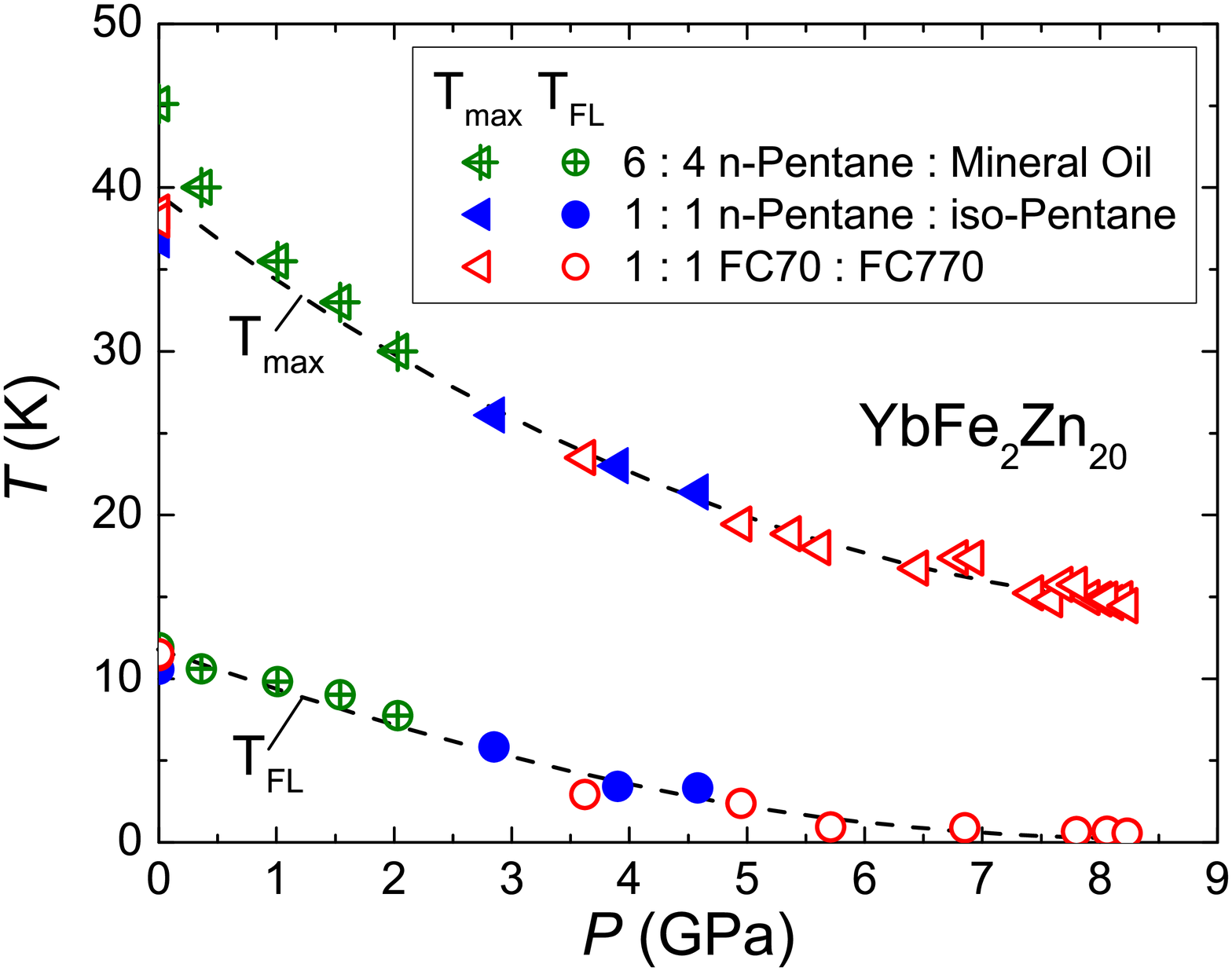}
\caption{(Color Online) \Tmax and \Tfl as they evolve with pressure.  The dashed lines are guides for the eye.}
\label{ZnTmax}
\end{figure}

The suppressions of \Tmax and \Tfl with pressure for \znfe are shown in Fig.~\ref{ZnTmax}.  If the minimum of \Tmax is indeed an indicator that the system will manifest magnetic ordering,\cite{Matsubayashi-JPCS10} then for \znfef , the critical pressure may well be at a slightly higher pressure than the maximum achieved in this study.  Furthermore, the Fermi-liquid state is almost completely suppressed by 8.23~GPa, another indicator that the system should be near its critical pressure value.  

\begin{figure}[!htbp] \centering
	%\begin{overpic}[width=80mm,grid]{ZnFL_PD.eps}
	\begin{overpic}[width=80mm]{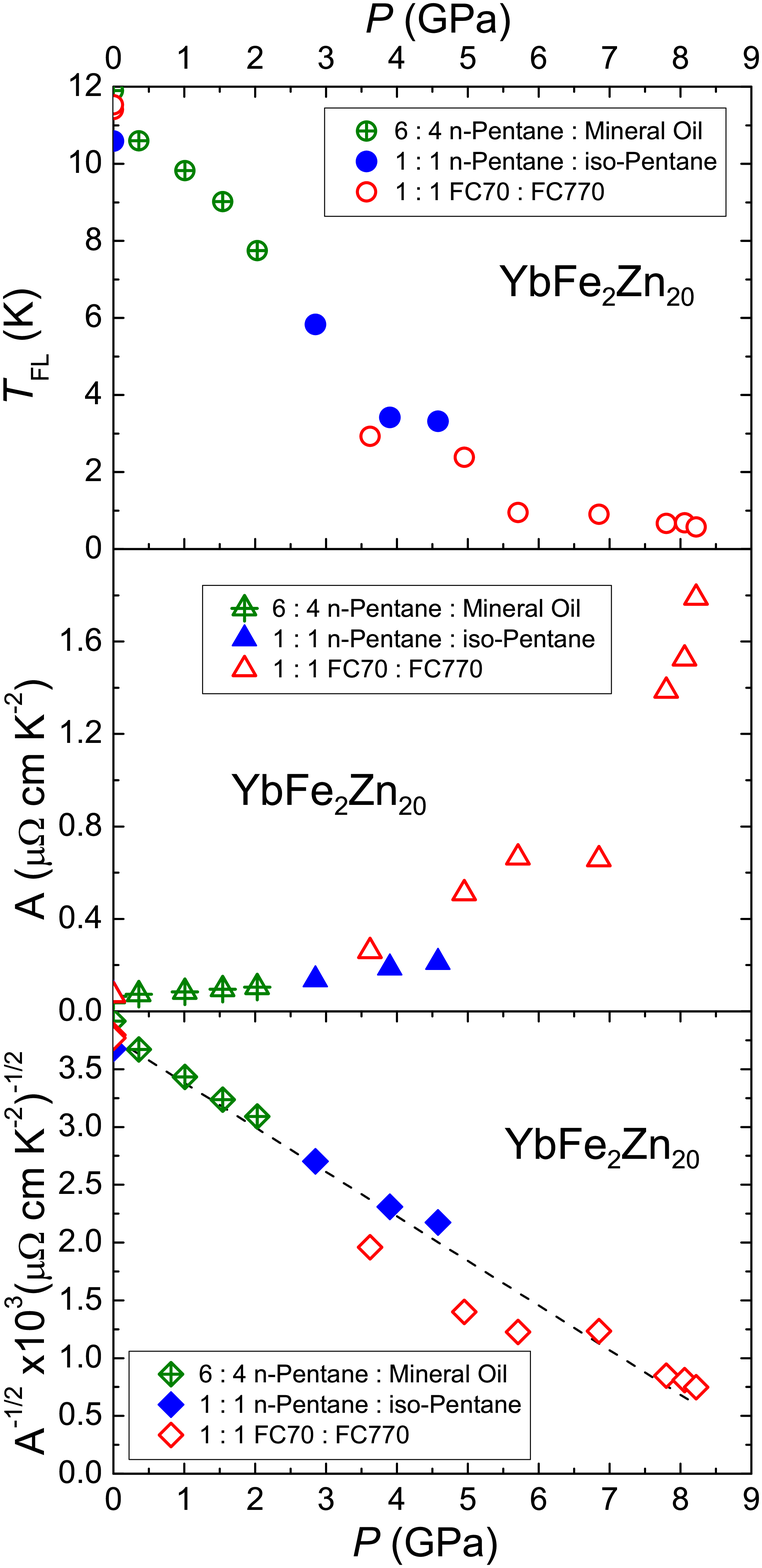}
		\put(9,84){(a)}
		\put(9,61){(b)}
		\put(9,28){(c)}
	\end{overpic}
	\caption{(Color Online) (a)~Evolution of \Tfl as pressure is applied.  By 8.23~GPa, \Tfl is only as high as $\sim$~0.6~K.  Crossed, closed, and open symbols are results from the piston cylinder cell, the mBAC with the \Pen mixture and mBAC with the \FC mixture, respectively.  (b)~\Tsq coefficient ($\rho\mathrm{(}T\mathrm{)}=\rho_0+AT^2$) which diverges as pressure is increased.  (c)~$A^{-1/2}$ as it progresses with pressure.  The dashed line is a linear fit to the data with $A^{-1/2}$ reaching zero near 9.8~GPa.}
	\label{ZnFL_PD}
\end{figure}

The influence of pressure on the Fermi-liquid behavior of \znfe is shown more clearly in Fig.~\ref{ZnFL_PD}.  In Fig.~\ref{ZnFL_PD}(a), the spread of \Tfl values seen at ambient pressure is small compared to the overall change in \Tfl over the 8~GPa pressure range.  Although \Tfl has not been driven to zero by our maximum pressure, extrapolation of \Tflf ($P$) suggests that this will occur at higher pressures in the range of 9-10~GPa.  The plot of the \Tsq coefficient, $A$, as a function of pressure (Fig.~\ref{ZnFL_PD}(b)) shows divergent behavior with pressure, consistent with the diminishing \Tfl and closeness to a quantum critical point.  

The divergence of $A\mathrm{(}P\mathrm{)}$ is very clearly quantified in Fig.~\ref{ZnFL_PD}(c) where the progression of $A^{-1/2}$ with pressure is shown.  A linear fit to the data, denoted by the black dashed line, shows that $A^{-1/2}$ will go to zero (indicating an infinitely large $A$ value) at $\sim 9.8$~GPa.  These data indicates that $A\propto \mathrm{(}P-P_{\mathrm{c}}\mathrm{)}^{-2}$ with $P_{\mathrm{c}}\simeq 9.8$~GPa.  It should be noted that data for $\rho_0\mathrm{(}P\mathrm{)}$ (not shown) does gradually increase from $P=0$ to 8.23~GPa, effectively doubling in value from ambient to highest pressure.

At our highest pressures, $A$ increases rapidly indicating a proximity to a quantum critical point  near $P_{\mathrm{c}}=9.8$~GPa.  As pressure increases from atmospheric pressure to $\sim 8$~GPa, we anticipate that the low temperature electronic specific heat, $\gamma$, increases.  By using the Kadowaki-Woods plot in Fig.~4 of Ref.~\onlinecite{Torikachvili-PNAS07}, we can predict an increase in $\gamma$ from 520~mJ/mol~K$^2$ to $\sim 2000$~mJ/mol~K$^2$ or greater at 8.23~GPa depending on whether all crystalline electric field split levels are comparable or lower than \Tkf .

For $P>P_{\mathrm{c}}\sim 9.8$~GPa, long-range magnetic ordering is expected. Such behavior was speculated to be seen with YbCo$_{2}$Zn$_{20}$, where at \Pc $\approx 1.0$~GPa, a small anomaly in the resistivity appeared near 0.15~K.\cite{Saiga-JPSJ08}  The large difference in the pressure needed to drive these systems to a quantum critical point is attributed to their large difference in ambient pressure Kondo temperatures (\Tk =~1.5~K for YbCo$_{2}$Zn$_{20}$ and \Tk =~33~K for \znfef ).\cite{Torikachvili-PNAS07,Matsubayashi-JPCS09}  This is also reflected in the value of \Pc for YbRh$_{2}$Zn$_{20}$ and YbIr$_{2}$Zn$_{20}$ which is suggested to be 3~GPa, where \Tk =~16 and 21~K, respectively.\cite{Torikachvili-PNAS07,Matsubayashi-JPCS09}  To estimate the hypothetical magnetic ordering temperature for trivalent Yb in \znfe and YbCo$_{2}$Zn$_{20}$, the ordering temperature for the Gd counterparts, GdFe$_{2}$Zn$_{20}$ and GdCo$_{2}$Zn$_{20}$, can be scaled by the de~Gennes factor.  The de~Gennes factor correlates the magnetic ordering temperature with the strength of the interaction between the local rare earth moment and the polarized conduction electrons.\cite{Jia-NP07,Jia-PRB09}  Using this method, YbFe$_{2}$Zn$_{20}$ and YbCo$_{2}$Zn$_{20}$ would be expected to order at 1.75 and 0.12~K, respectively.  For YbCo$_{2}$Zn$_{20}$, this is relatively close to the value of 0.15~K measured under pressure,\cite{Saiga-JPSJ08} suggesting that magnetic ordering in \znfe might be found as high as $\sim 1$~K for $P\gtrsim 9.8$~GPa.  

Furthermore, RFe$_2$Zn$_{20}$ compounds, for R~=~Gd-Tm, manifest ferromagnetic ordering with the Curie temperature, \TCf , for Gd as high as 86~K.\cite{Jia-NP07,Jia-PRB09}  The enhanced \TCf 's of these compounds are thought to be associated with the near Stoner nature of YFe$_2$Zn$_{20}$ and LuFe$_2$Zn$_{20}$.\cite{Jia-NP07,Jia-Thesis08,Jia-PRB09}  This then raises the possibility that the magnetic ordering that evolves with pressure from the Fermi-liquid state of \znfe might be more complex than simple antiferromagnetic order.

\begin{figure}[!htbp]
\centering
\includegraphics[width=80mm]{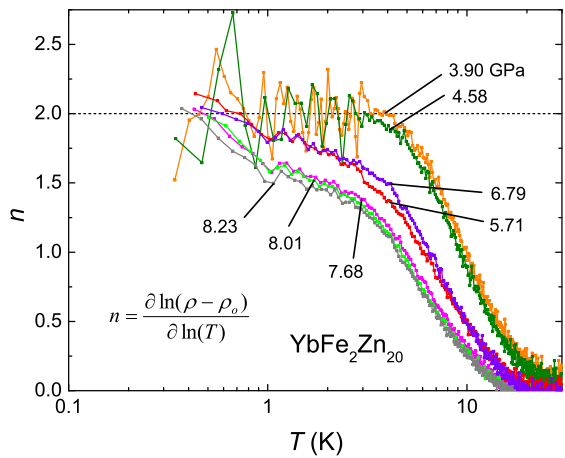}
\caption{(Color Online) Power law behavior for selected resistivity curves on a semilog scale.  Data for $T<3$~K are from measurements with a $^3$He cryostat and data for $T>3$~K are from measurements with the PPMS.  The black dotted line is a guide for the eye and indicates a power of $n=2$ for $\rho\mathrm{(}T\mathrm{)}=\rho_0+AT^n$.}
\label{ZnPL}
\end{figure}

To further investigate the possibility that a QCP is being approached, an analysis of the progression of the power law behavior at low temperatures and high pressures was done.  Figure~\ref{ZnPL} shows the results for pressures where measurements down to $\sim 0.3-0.4$~K were taken.  Data from base temperature to 3~K are results from measurements in the $^3$He cryostat and above 3~K, results from measurements in the PPMS.  In this analysis, the lowest temperature $\rho_{0}$ from a $T^2$ fit was used for both sets of measurements.  

As expected, low pressure measurements show $n=2$ at base temperatures across a decade of temperature.  The large noise in the low temperature data for 3.90 and 4.58~GPa is restricted to one particular packing of the pressure cell (with higher noise) as the higher pressure $n$ values (from different pressure cell packings) do not suffer from such large noise.  For pressures above 4.58~GPa, $n=2$ for a short range at low temperatures then gradually decreases as temperature increases.  For the highest pressures where $P\geq 5.71$~GPa, $n=2$ for only the lowest temperatures and instead shows $n$ close to 1.5 for a significant temperature range. Since at our highest pressure, 8.23~GPa, is still below the expected $P_{\mathrm{c}}\approx 9.8$~GPa, it is still possible that by 9.8~GPa, the system will manifest an $n=1.5$ power law behavior over a wider temperature range.  If AFM order does arise at the critical pressure, this would be consistent with the results for spin fluctuation theories by Hertz and Millis as well as Moriya\cite{Hertz-PRB76,Millis-PRB93,Moriya-JPSJ95} where an $n$~=~1.5 power law is predicted for a 3D AFM system.  On the other hand, if ferromagnetic ordering occurs, $n\approx 1.33$ or 1.67 is predicted by Moriya\cite{Moriya-JPSJ95} for 2D or 3D systems, respectively.  

Another possibility is that instead of a QCP, we may be approaching a quantum critical region where neither Fermi-liquid behavior nor magnetic ordering exists and instead another exotic ground state may stabilize which is then followed by, at higher pressure, a magnetically ordered state.\cite{Budko-PRB04,Budko-PRB05,Budko-PRB05a,Custers-PRL10,Schmiedeshoff-PRB11,Mun-PRB13}  It is evident that higher pressure measurements are necessary to further explore these possibilities.

%-------------------------------------------------------------------------------
%-------------------------------------------------------------------------------

\section{Conclusion}
The resistivity of \znfe was measured under pressure up to 8.23~GPa and down to temperatures of almost 0.3~K.  Increasing pressure drives the characteristic Kondo temperature, \Tkf , to lower temperatures and diminishes the range of Fermi-liquid behavior.  The dramatic enhancement of $A$ as pressure increases to 8.23~GPa suggests a close proximity of \znfef , at our highest pressure, to a quantum critical point where the system may develop a new magnetic ground state.  Although not reached in this study, the critical pressure for \znfe can be inferred from $A\mathrm{(}P\mathrm{)}\propto \mathrm{(}P-P_{\mathrm{c}}\mathrm{)}^{-2}$ to be $P_{\mathrm{c}}\simeq 9.8$~GPa.  

\begin{acknowledgments}
The authors would like to thank S.~Jia and E.~Mun for growing the \znfe samples used in this study and H.~Hodovanets, V.~Taufour, for technical assistance and fruitful discussions.  This work was carried out at Ames Laboratory, US DOE, under Contract \# DE-AC02-07CH11358 (S.K.K., S.L.B. and P.C.C.). Part of this work was performed at the Iowa State University and supported by the AFOSR-MURI Grant \# FA9550-09-1-0603 and also by the National Science Foundation under Grant \# DMR-0805335 (M.S.T.). S.L.B. acknowledges partial support from the State of Iowa through Iowa State University.  
\end{acknowledgments}

%\clearpage

\bibliography{C:/Users/skim/Documents/Papers/Masterbib}

\begin{thebibliography}{42}
\expandafter\ifx\csname natexlab\endcsname\relax\def\natexlab#1{#1}\fi
\expandafter\ifx\csname bibnamefont\endcsname\relax
  \def\bibnamefont#1{#1}\fi
\expandafter\ifx\csname bibfnamefont\endcsname\relax
  \def\bibfnamefont#1{#1}\fi
\expandafter\ifx\csname citenamefont\endcsname\relax
  \def\citenamefont#1{#1}\fi
\expandafter\ifx\csname url\endcsname\relax
  \def\url#1{\texttt{#1}}\fi
\expandafter\ifx\csname urlprefix\endcsname\relax\def\urlprefix{URL }\fi
\providecommand{\bibinfo}[2]{#2}
\providecommand{\eprint}[2][]{\url{#2}}

\bibitem[{\citenamefont{Mott}(1990)}]{Mott-MIT90}
\bibinfo{author}{\bibfnamefont{N.}~\bibnamefont{Mott}},
  \emph{\bibinfo{title}{{M}etal-{I}nsulator {T}ransitions: {S}econd {E}dition}}
  (\bibinfo{publisher}{Taylor \& Francis}, \bibinfo{year}{1990}).

\bibitem[{\citenamefont{Bennemann and Ketterson}(2008)}]{Bennemann-08}
\bibinfo{author}{\bibfnamefont{K.}~\bibnamefont{Bennemann}} \bibnamefont{and}
  \bibinfo{author}{\bibfnamefont{J.}~\bibnamefont{Ketterson}},
  \emph{\bibinfo{title}{{S}uperconductivity: {N}ovel superconductors}}
  (\bibinfo{publisher}{Springer-Verlag Berlin Heidelberg},
  \bibinfo{year}{2008}).

\bibitem[{\citenamefont{Norman}(2011)}]{Norman-SM11}
\bibinfo{author}{\bibfnamefont{M.~R.} \bibnamefont{Norman}},
  \bibinfo{journal}{Science} \textbf{\bibinfo{volume}{332}},
  \bibinfo{pages}{196} (\bibinfo{year}{2011}).

\bibitem[{\citenamefont{Stewart}(1984)}]{Stewart-RMP84}
\bibinfo{author}{\bibfnamefont{G.~R.} \bibnamefont{Stewart}},
  \bibinfo{journal}{Rev. Mod. Phys.} \textbf{\bibinfo{volume}{56}},
  \bibinfo{pages}{755} (\bibinfo{year}{1984}).

\bibitem[{\citenamefont{Stewart}(2001)}]{Stewart-RMP01}
\bibinfo{author}{\bibfnamefont{G.~R.} \bibnamefont{Stewart}},
  \bibinfo{journal}{Rev. Mod. Phys.} \textbf{\bibinfo{volume}{73}},
  \bibinfo{pages}{797} (\bibinfo{year}{2001}).

\bibitem[{\citenamefont{Stewart}(2006)}]{Stewart-RMP06}
\bibinfo{author}{\bibfnamefont{G.~R.} \bibnamefont{Stewart}},
  \bibinfo{journal}{Rev. Mod. Phys.} \textbf{\bibinfo{volume}{78}},
  \bibinfo{pages}{743} (\bibinfo{year}{2006}).

\bibitem[{\citenamefont{Torikachvili et~al.}(2007)\citenamefont{Torikachvili,
  Jia, Mun, Hannahs, Black, Neils, Martien, Bud'ko, and
  Canfield}}]{Torikachvili-PNAS07}
\bibinfo{author}{\bibfnamefont{M.~S.} \bibnamefont{Torikachvili}},
  \bibinfo{author}{\bibfnamefont{S.}~\bibnamefont{Jia}},
  \bibinfo{author}{\bibfnamefont{E.~D.} \bibnamefont{Mun}},
  \bibinfo{author}{\bibfnamefont{S.~T.} \bibnamefont{Hannahs}},
  \bibinfo{author}{\bibfnamefont{R.~C.} \bibnamefont{Black}},
  \bibinfo{author}{\bibfnamefont{W.~K.} \bibnamefont{Neils}},
  \bibinfo{author}{\bibfnamefont{D.}~\bibnamefont{Martien}},
  \bibinfo{author}{\bibfnamefont{S.~L.} \bibnamefont{Bud'ko}},
  \bibnamefont{and} \bibinfo{author}{\bibfnamefont{P.~C.}
  \bibnamefont{Canfield}}, \bibinfo{journal}{Proc. Natl. Acad. Sci.}
  \textbf{\bibinfo{volume}{104}}, \bibinfo{pages}{9960} (\bibinfo{year}{2007}).

\bibitem[{\citenamefont{Canfield et~al.}(2008)\citenamefont{Canfield, Jia, Mun,
  Bud'ko, Samolyuk, and Torikachvili}}]{Canfield-PB08}
\bibinfo{author}{\bibfnamefont{P.}~\bibnamefont{Canfield}},
  \bibinfo{author}{\bibfnamefont{S.}~\bibnamefont{Jia}},
  \bibinfo{author}{\bibfnamefont{E.}~\bibnamefont{Mun}},
  \bibinfo{author}{\bibfnamefont{S.}~\bibnamefont{Bud'ko}},
  \bibinfo{author}{\bibfnamefont{G.}~\bibnamefont{Samolyuk}}, \bibnamefont{and}
  \bibinfo{author}{\bibfnamefont{M.}~\bibnamefont{Torikachvili}},
  \bibinfo{journal}{Physica B} \textbf{\bibinfo{volume}{403}},
  \bibinfo{pages}{844} (\bibinfo{year}{2008}).

\bibitem[{\citenamefont{Jia et~al.}(2009)\citenamefont{Jia, Ni, Bud'ko, and
  Canfield}}]{Jia-PRB09}
\bibinfo{author}{\bibfnamefont{S.}~\bibnamefont{Jia}},
  \bibinfo{author}{\bibfnamefont{N.}~\bibnamefont{Ni}},
  \bibinfo{author}{\bibfnamefont{S.~L.} \bibnamefont{Bud'ko}},
  \bibnamefont{and} \bibinfo{author}{\bibfnamefont{P.~C.}
  \bibnamefont{Canfield}}, \bibinfo{journal}{Phys. Rev. B}
  \textbf{\bibinfo{volume}{80}}, \bibinfo{pages}{104403}
  (\bibinfo{year}{2009}).

\bibitem[{\citenamefont{Jia}(2008)}]{Jia-Thesis08}
\bibinfo{author}{\bibfnamefont{S.}~\bibnamefont{Jia}}, Ph.D. thesis,
  \bibinfo{school}{Iowa State University} (\bibinfo{year}{2008}).

\bibitem[{\citenamefont{Mun et~al.}(2012)\citenamefont{Mun, Jia, Bud'ko, and
  Canfield}}]{Mun-PRB12}
\bibinfo{author}{\bibfnamefont{E.~D.} \bibnamefont{Mun}},
  \bibinfo{author}{\bibfnamefont{S.}~\bibnamefont{Jia}},
  \bibinfo{author}{\bibfnamefont{S.~L.} \bibnamefont{Bud'ko}},
  \bibnamefont{and} \bibinfo{author}{\bibfnamefont{P.~C.}
  \bibnamefont{Canfield}}, \bibinfo{journal}{Phys. Rev. B}
  \textbf{\bibinfo{volume}{86}}, \bibinfo{pages}{115110}
  (\bibinfo{year}{2012}).

\bibitem[{\citenamefont{Coleman}(2007)}]{Coleman-07}
\bibinfo{author}{\bibfnamefont{P.}~\bibnamefont{Coleman}},
  \emph{\bibinfo{title}{{H}eavy {F}ermions: {E}lectrons at the {E}dge of
  {M}agnetism}} (\bibinfo{publisher}{John Wiley \& Sons, Ltd},
  \bibinfo{year}{2007}), chap.~\bibinfo{chapter}{2}, pp.
  \bibinfo{pages}{95--148}.

\bibitem[{\citenamefont{Doniach}(1977)}]{Doniach-PB77}
\bibinfo{author}{\bibfnamefont{S.}~\bibnamefont{Doniach}},
  \bibinfo{journal}{Physica B+C} \textbf{\bibinfo{volume}{91}},
  \bibinfo{pages}{231 } (\bibinfo{year}{1977}).

\bibitem[{\citenamefont{Saiga et~al.}(2008)\citenamefont{Saiga, Matsubayashi,
  Fujiwara, Kosaka, Katano, Hedo, Matsumoto, and Uwatoko}}]{Saiga-JPSJ08}
\bibinfo{author}{\bibfnamefont{Y.}~\bibnamefont{Saiga}},
  \bibinfo{author}{\bibfnamefont{K.}~\bibnamefont{Matsubayashi}},
  \bibinfo{author}{\bibfnamefont{T.}~\bibnamefont{Fujiwara}},
  \bibinfo{author}{\bibfnamefont{M.}~\bibnamefont{Kosaka}},
  \bibinfo{author}{\bibfnamefont{S.}~\bibnamefont{Katano}},
  \bibinfo{author}{\bibfnamefont{M.}~\bibnamefont{Hedo}},
  \bibinfo{author}{\bibfnamefont{T.}~\bibnamefont{Matsumoto}},
  \bibnamefont{and} \bibinfo{author}{\bibfnamefont{Y.}~\bibnamefont{Uwatoko}},
  \bibinfo{journal}{J. Phys. Soc. Jpn.} \textbf{\bibinfo{volume}{77}},
  \bibinfo{pages}{053710} (\bibinfo{year}{2008}).

\bibitem[{\citenamefont{Matsubayashi et~al.}(2010)\citenamefont{Matsubayashi,
  Saiga, Matsumoto, and Uwatoko}}]{Matsubayashi-JPCS10}
\bibinfo{author}{\bibfnamefont{K.}~\bibnamefont{Matsubayashi}},
  \bibinfo{author}{\bibfnamefont{Y.}~\bibnamefont{Saiga}},
  \bibinfo{author}{\bibfnamefont{T.}~\bibnamefont{Matsumoto}},
  \bibnamefont{and} \bibinfo{author}{\bibfnamefont{Y.}~\bibnamefont{Uwatoko}},
  \bibinfo{journal}{J. Phys.: Conf. Ser.} \textbf{\bibinfo{volume}{200}},
  \bibinfo{pages}{012112} (\bibinfo{year}{2010}).

\bibitem[{\citenamefont{Matsubayashi et~al.}(2009)\citenamefont{Matsubayashi,
  Saiga, Matsumoto, and Uwatoko}}]{Matsubayashi-JPCS09}
\bibinfo{author}{\bibfnamefont{K.}~\bibnamefont{Matsubayashi}},
  \bibinfo{author}{\bibfnamefont{Y.}~\bibnamefont{Saiga}},
  \bibinfo{author}{\bibfnamefont{T.}~\bibnamefont{Matsumoto}},
  \bibnamefont{and} \bibinfo{author}{\bibfnamefont{Y.}~\bibnamefont{Uwatoko}},
  \bibinfo{journal}{J. Phys.: Conf. Ser.} \textbf{\bibinfo{volume}{150}},
  \bibinfo{pages}{042117} (\bibinfo{year}{2009}).

\bibitem[{\citenamefont{Jia et~al.}(2007)\citenamefont{Jia, Bud'ko, Samolyuk,
  and Canfield}}]{Jia-NP07}
\bibinfo{author}{\bibfnamefont{S.}~\bibnamefont{Jia}},
  \bibinfo{author}{\bibfnamefont{S.~L.} \bibnamefont{Bud'ko}},
  \bibinfo{author}{\bibfnamefont{G.~D.} \bibnamefont{Samolyuk}},
  \bibnamefont{and} \bibinfo{author}{\bibfnamefont{P.~C.}
  \bibnamefont{Canfield}}, \bibinfo{journal}{Nat. Phys.}
  \textbf{\bibinfo{volume}{3}}, \bibinfo{pages}{334} (\bibinfo{year}{2007}).

\bibitem[{\citenamefont{Canfield}(2010)}]{Canfield-Euroschool10}
\bibinfo{author}{\bibfnamefont{P.~C.} \bibnamefont{Canfield}},
  \emph{\bibinfo{title}{{S}olution growth of intermetallic single crystals: a
  beginner's guide}} (\bibinfo{publisher}{World Scientific},
  \bibinfo{year}{2010}), vol.~\bibinfo{volume}{2} of
  \emph{\bibinfo{series}{Book Series on Complex Metallic Alloys}},
  chap.~\bibinfo{chapter}{2}, pp. \bibinfo{pages}{93--111}.

\bibitem[{\citenamefont{Torikachvili et~al.}(2008)\citenamefont{Torikachvili,
  Bud'ko, Ni, and Canfield}}]{Torikachvili-PRB08}
\bibinfo{author}{\bibfnamefont{M.~S.} \bibnamefont{Torikachvili}},
  \bibinfo{author}{\bibfnamefont{S.~L.} \bibnamefont{Bud'ko}},
  \bibinfo{author}{\bibfnamefont{N.}~\bibnamefont{Ni}}, \bibnamefont{and}
  \bibinfo{author}{\bibfnamefont{P.~C.} \bibnamefont{Canfield}},
  \bibinfo{journal}{Phys. Rev. B} \textbf{\bibinfo{volume}{78}},
  \bibinfo{pages}{104527} (\bibinfo{year}{2008}).

\bibitem[{\citenamefont{Torikachvili et~al.}(2009)\citenamefont{Torikachvili,
  Bud'ko, Ni, Canfield, and Hannahs}}]{Torikachvili-PRB09}
\bibinfo{author}{\bibfnamefont{M.~S.} \bibnamefont{Torikachvili}},
  \bibinfo{author}{\bibfnamefont{S.~L.} \bibnamefont{Bud'ko}},
  \bibinfo{author}{\bibfnamefont{N.}~\bibnamefont{Ni}},
  \bibinfo{author}{\bibfnamefont{P.~C.} \bibnamefont{Canfield}},
  \bibnamefont{and} \bibinfo{author}{\bibfnamefont{S.~T.}
  \bibnamefont{Hannahs}}, \bibinfo{journal}{Phys. Rev. B}
  \textbf{\bibinfo{volume}{80}}, \bibinfo{pages}{014521}
  (\bibinfo{year}{2009}).

\bibitem[{\citenamefont{Colombier and Braithwaite}(2007)}]{Colombier-RSI07}
\bibinfo{author}{\bibfnamefont{E.}~\bibnamefont{Colombier}} \bibnamefont{and}
  \bibinfo{author}{\bibfnamefont{D.}~\bibnamefont{Braithwaite}},
  \bibinfo{journal}{Rev. Sci. Instrum.} \textbf{\bibinfo{volume}{78}},
  \bibinfo{eid}{093903} (\bibinfo{year}{2007}).

\bibitem[{\citenamefont{Eiling and Schilling}(1981)}]{Eiling-JPF81}
\bibinfo{author}{\bibfnamefont{A.}~\bibnamefont{Eiling}} \bibnamefont{and}
  \bibinfo{author}{\bibfnamefont{J.~S.} \bibnamefont{Schilling}},
  \bibinfo{journal}{J. Phys. F} \textbf{\bibinfo{volume}{11}},
  \bibinfo{pages}{623} (\bibinfo{year}{1981}).

\bibitem[{\citenamefont{Piermarini et~al.}(1973)\citenamefont{Piermarini,
  Block, and Barnett}}]{Piermarini-JAP73}
\bibinfo{author}{\bibfnamefont{G.~J.} \bibnamefont{Piermarini}},
  \bibinfo{author}{\bibfnamefont{S.}~\bibnamefont{Block}}, \bibnamefont{and}
  \bibinfo{author}{\bibfnamefont{J.}~\bibnamefont{Barnett}},
  \bibinfo{journal}{J. Appl. Phys.} \textbf{\bibinfo{volume}{44}},
  \bibinfo{pages}{5377} (\bibinfo{year}{1973}).

\bibitem[{\citenamefont{Klotz et~al.}(2009)\citenamefont{Klotz, Chervin,
  Munsch, and Marchand}}]{Klotz-JPDAP09}
\bibinfo{author}{\bibfnamefont{S.}~\bibnamefont{Klotz}},
  \bibinfo{author}{\bibfnamefont{J.-C.} \bibnamefont{Chervin}},
  \bibinfo{author}{\bibfnamefont{P.}~\bibnamefont{Munsch}}, \bibnamefont{and}
  \bibinfo{author}{\bibfnamefont{G.~L.} \bibnamefont{Marchand}},
  \bibinfo{journal}{J. Phys. D} \textbf{\bibinfo{volume}{42}},
  \bibinfo{pages}{075413} (\bibinfo{year}{2009}).

\bibitem[{\citenamefont{Varga et~al.}(2003)\citenamefont{Varga, Wilkinson, and
  Angel}}]{Varga-RSI03}
\bibinfo{author}{\bibfnamefont{T.}~\bibnamefont{Varga}},
  \bibinfo{author}{\bibfnamefont{A.~P.} \bibnamefont{Wilkinson}},
  \bibnamefont{and} \bibinfo{author}{\bibfnamefont{R.~J.} \bibnamefont{Angel}},
  \bibinfo{journal}{Review of Scientific Instruments}
  \textbf{\bibinfo{volume}{74}}, \bibinfo{pages}{4564} (\bibinfo{year}{2003}).

\bibitem[{\citenamefont{Sidorov and Sadykov}(2005)}]{Sidorov-JPCM05}
\bibinfo{author}{\bibfnamefont{V.~A.} \bibnamefont{Sidorov}} \bibnamefont{and}
  \bibinfo{author}{\bibfnamefont{R.~A.} \bibnamefont{Sadykov}},
  \bibinfo{journal}{J. Phys.: Condens. Matter} \textbf{\bibinfo{volume}{17}},
  \bibinfo{pages}{S3005} (\bibinfo{year}{2005}).

\bibitem[{\citenamefont{Alami-Yadri et~al.}(1998)\citenamefont{Alami-Yadri,
  Wilhelm, and Jaccard}}]{Alami-SSC98}
\bibinfo{author}{\bibfnamefont{K.}~\bibnamefont{Alami-Yadri}},
  \bibinfo{author}{\bibfnamefont{H.}~\bibnamefont{Wilhelm}}, \bibnamefont{and}
  \bibinfo{author}{\bibfnamefont{D.}~\bibnamefont{Jaccard}},
  \bibinfo{journal}{Solid State Commun.} \textbf{\bibinfo{volume}{108}},
  \bibinfo{pages}{279} (\bibinfo{year}{1998}).

\bibitem[{\citenamefont{Knebel et~al.}(2001)\citenamefont{Knebel, Braithwaite,
  Lapertot, Canfield, and Flouquet}}]{Knebel-JPCM01}
\bibinfo{author}{\bibfnamefont{G.}~\bibnamefont{Knebel}},
  \bibinfo{author}{\bibfnamefont{D.}~\bibnamefont{Braithwaite}},
  \bibinfo{author}{\bibfnamefont{G.}~\bibnamefont{Lapertot}},
  \bibinfo{author}{\bibfnamefont{P.~C.} \bibnamefont{Canfield}},
  \bibnamefont{and} \bibinfo{author}{\bibfnamefont{J.}~\bibnamefont{Flouquet}},
  \bibinfo{journal}{J. Phys.: Condens. Matter} \textbf{\bibinfo{volume}{13}},
  \bibinfo{pages}{10935} (\bibinfo{year}{2001}).

\bibitem[{\citenamefont{Trovarelli et~al.}(2002)\citenamefont{Trovarelli,
  Custers, Gegenwart, Geibel, Hinze, Mederle, Sparn, and
  Steglich}}]{Trovarelli-PBCM02}
\bibinfo{author}{\bibfnamefont{O.}~\bibnamefont{Trovarelli}},
  \bibinfo{author}{\bibfnamefont{J.}~\bibnamefont{Custers}},
  \bibinfo{author}{\bibfnamefont{P.}~\bibnamefont{Gegenwart}},
  \bibinfo{author}{\bibfnamefont{C.}~\bibnamefont{Geibel}},
  \bibinfo{author}{\bibfnamefont{P.}~\bibnamefont{Hinze}},
  \bibinfo{author}{\bibfnamefont{S.}~\bibnamefont{Mederle}},
  \bibinfo{author}{\bibfnamefont{G.}~\bibnamefont{Sparn}}, \bibnamefont{and}
  \bibinfo{author}{\bibfnamefont{F.}~\bibnamefont{Steglich}},
  \bibinfo{journal}{Physica B} \textbf{\bibinfo{volume}{312-313}},
  \bibinfo{pages}{401} (\bibinfo{year}{2002}).

\bibitem[{\citenamefont{Winkelmann et~al.}(1998)\citenamefont{Winkelmann,
  Abd-Elmeguid, Micklitz, Sanchez, Geibel, and Steglich}}]{Winkelmann-PRL98}
\bibinfo{author}{\bibfnamefont{H.}~\bibnamefont{Winkelmann}},
  \bibinfo{author}{\bibfnamefont{M.~M.} \bibnamefont{Abd-Elmeguid}},
  \bibinfo{author}{\bibfnamefont{H.}~\bibnamefont{Micklitz}},
  \bibinfo{author}{\bibfnamefont{J.~P.} \bibnamefont{Sanchez}},
  \bibinfo{author}{\bibfnamefont{C.}~\bibnamefont{Geibel}}, \bibnamefont{and}
  \bibinfo{author}{\bibfnamefont{F.}~\bibnamefont{Steglich}},
  \bibinfo{journal}{Phys. Rev. Lett.} \textbf{\bibinfo{volume}{81}},
  \bibinfo{pages}{4947} (\bibinfo{year}{1998}).

\bibitem[{\citenamefont{Yoshimori and Kasai}(1983)}]{Yoshimori-JMMM83}
\bibinfo{author}{\bibfnamefont{A.}~\bibnamefont{Yoshimori}} \bibnamefont{and}
  \bibinfo{author}{\bibfnamefont{H.}~\bibnamefont{Kasai}}, \bibinfo{journal}{J.
  Magn. Magn. Mater.} \textbf{\bibinfo{volume}{31-34, Part 1}},
  \bibinfo{pages}{475} (\bibinfo{year}{1983}).

\bibitem[{\citenamefont{Coleman}(1987)}]{Coleman-JMMM87}
\bibinfo{author}{\bibfnamefont{P.}~\bibnamefont{Coleman}}, \bibinfo{journal}{J.
  Magn. Magn. Mater.} \textbf{\bibinfo{volume}{63-64}}, \bibinfo{pages}{245}
  (\bibinfo{year}{1987}).

\bibitem[{\citenamefont{Bauer et~al.}(1993)\citenamefont{Bauer, Hauser, Gratz,
  Payer, Oomi, and Kagayama}}]{Bauer-PRB93}
\bibinfo{author}{\bibfnamefont{E.}~\bibnamefont{Bauer}},
  \bibinfo{author}{\bibfnamefont{R.}~\bibnamefont{Hauser}},
  \bibinfo{author}{\bibfnamefont{E.}~\bibnamefont{Gratz}},
  \bibinfo{author}{\bibfnamefont{K.}~\bibnamefont{Payer}},
  \bibinfo{author}{\bibfnamefont{G.}~\bibnamefont{Oomi}}, \bibnamefont{and}
  \bibinfo{author}{\bibfnamefont{T.}~\bibnamefont{Kagayama}},
  \bibinfo{journal}{Phys. Rev. B} \textbf{\bibinfo{volume}{48}},
  \bibinfo{pages}{15873} (\bibinfo{year}{1993}).

\bibitem[{\citenamefont{Hertz}(1976)}]{Hertz-PRB76}
\bibinfo{author}{\bibfnamefont{J.~A.} \bibnamefont{Hertz}},
  \bibinfo{journal}{Phys. Rev. B} \textbf{\bibinfo{volume}{14}},
  \bibinfo{pages}{1165} (\bibinfo{year}{1976}).

\bibitem[{\citenamefont{Millis}(1993)}]{Millis-PRB93}
\bibinfo{author}{\bibfnamefont{A.~J.} \bibnamefont{Millis}},
  \bibinfo{journal}{Phys. Rev. B} \textbf{\bibinfo{volume}{48}},
  \bibinfo{pages}{7183} (\bibinfo{year}{1993}).

\bibitem[{\citenamefont{Moriya and Takimoto}(1995)}]{Moriya-JPSJ95}
\bibinfo{author}{\bibfnamefont{T.}~\bibnamefont{Moriya}} \bibnamefont{and}
  \bibinfo{author}{\bibfnamefont{T.}~\bibnamefont{Takimoto}},
  \bibinfo{journal}{J. Phys. Soc. Jpn.} \textbf{\bibinfo{volume}{64}},
  \bibinfo{pages}{960} (\bibinfo{year}{1995}).

\bibitem[{\citenamefont{Bud'ko et~al.}(2004)\citenamefont{Bud'ko, Morosan, and
  Canfield}}]{Budko-PRB04}
\bibinfo{author}{\bibfnamefont{S.~L.} \bibnamefont{Bud'ko}},
  \bibinfo{author}{\bibfnamefont{E.}~\bibnamefont{Morosan}}, \bibnamefont{and}
  \bibinfo{author}{\bibfnamefont{P.~C.} \bibnamefont{Canfield}},
  \bibinfo{journal}{Phys. Rev. B} \textbf{\bibinfo{volume}{69}},
  \bibinfo{pages}{014415} (\bibinfo{year}{2004}).

\bibitem[{\citenamefont{Bud'ko et~al.}(2005{\natexlab{a}})\citenamefont{Bud'ko,
  Morosan, and Canfield}}]{Budko-PRB05}
\bibinfo{author}{\bibfnamefont{S.~L.} \bibnamefont{Bud'ko}},
  \bibinfo{author}{\bibfnamefont{E.}~\bibnamefont{Morosan}}, \bibnamefont{and}
  \bibinfo{author}{\bibfnamefont{P.~C.} \bibnamefont{Canfield}},
  \bibinfo{journal}{Phys. Rev. B} \textbf{\bibinfo{volume}{71}},
  \bibinfo{pages}{054408} (\bibinfo{year}{2005}{\natexlab{a}}).

\bibitem[{\citenamefont{Bud'ko et~al.}(2005{\natexlab{b}})\citenamefont{Bud'ko,
  Zapf, Morosan, and Canfield}}]{Budko-PRB05a}
\bibinfo{author}{\bibfnamefont{S.~L.} \bibnamefont{Bud'ko}},
  \bibinfo{author}{\bibfnamefont{V.}~\bibnamefont{Zapf}},
  \bibinfo{author}{\bibfnamefont{E.}~\bibnamefont{Morosan}}, \bibnamefont{and}
  \bibinfo{author}{\bibfnamefont{P.~C.} \bibnamefont{Canfield}},
  \bibinfo{journal}{Phys. Rev. B} \textbf{\bibinfo{volume}{72}},
  \bibinfo{pages}{172413} (\bibinfo{year}{2005}{\natexlab{b}}).

\bibitem[{\citenamefont{Custers et~al.}(2010)\citenamefont{Custers, Gegenwart,
  Geibel, Steglich, Coleman, and Paschen}}]{Custers-PRL10}
\bibinfo{author}{\bibfnamefont{J.}~\bibnamefont{Custers}},
  \bibinfo{author}{\bibfnamefont{P.}~\bibnamefont{Gegenwart}},
  \bibinfo{author}{\bibfnamefont{C.}~\bibnamefont{Geibel}},
  \bibinfo{author}{\bibfnamefont{F.}~\bibnamefont{Steglich}},
  \bibinfo{author}{\bibfnamefont{P.}~\bibnamefont{Coleman}}, \bibnamefont{and}
  \bibinfo{author}{\bibfnamefont{S.}~\bibnamefont{Paschen}},
  \bibinfo{journal}{Phys. Rev. Lett.} \textbf{\bibinfo{volume}{104}},
  \bibinfo{pages}{186402} (\bibinfo{year}{2010}).

\bibitem[{\citenamefont{Schmiedeshoff et~al.}(2011)\citenamefont{Schmiedeshoff,
  Mun, Lounsbury, Tracy, Palm, Hannahs, Park, Murphy, Bud'ko, and
  Canfield}}]{Schmiedeshoff-PRB11}
\bibinfo{author}{\bibfnamefont{G.~M.} \bibnamefont{Schmiedeshoff}},
  \bibinfo{author}{\bibfnamefont{E.~D.} \bibnamefont{Mun}},
  \bibinfo{author}{\bibfnamefont{A.~W.} \bibnamefont{Lounsbury}},
  \bibinfo{author}{\bibfnamefont{S.~J.} \bibnamefont{Tracy}},
  \bibinfo{author}{\bibfnamefont{E.~C.} \bibnamefont{Palm}},
  \bibinfo{author}{\bibfnamefont{S.~T.} \bibnamefont{Hannahs}},
  \bibinfo{author}{\bibfnamefont{J.-H.} \bibnamefont{Park}},
  \bibinfo{author}{\bibfnamefont{T.~P.} \bibnamefont{Murphy}},
  \bibinfo{author}{\bibfnamefont{S.~L.} \bibnamefont{Bud'ko}},
  \bibnamefont{and} \bibinfo{author}{\bibfnamefont{P.~C.}
  \bibnamefont{Canfield}}, \bibinfo{journal}{Phys. Rev. B}
  \textbf{\bibinfo{volume}{83}}, \bibinfo{pages}{180408}
  (\bibinfo{year}{2011}).

\bibitem[{\citenamefont{Mun et~al.}(2013)\citenamefont{Mun, Bud'ko, Martin,
  Kim, Tanatar, Park, Murphy, Schmiedeshoff, Dilley, Prozorov
  et~al.}}]{Mun-PRB13}
\bibinfo{author}{\bibfnamefont{E.~D.} \bibnamefont{Mun}},
  \bibinfo{author}{\bibfnamefont{S.~L.} \bibnamefont{Bud'ko}},
  \bibinfo{author}{\bibfnamefont{C.}~\bibnamefont{Martin}},
  \bibinfo{author}{\bibfnamefont{H.}~\bibnamefont{Kim}},
  \bibinfo{author}{\bibfnamefont{M.~A.} \bibnamefont{Tanatar}},
  \bibinfo{author}{\bibfnamefont{J.-H.} \bibnamefont{Park}},
  \bibinfo{author}{\bibfnamefont{T.}~\bibnamefont{Murphy}},
  \bibinfo{author}{\bibfnamefont{G.~M.} \bibnamefont{Schmiedeshoff}},
  \bibinfo{author}{\bibfnamefont{N.}~\bibnamefont{Dilley}},
  \bibinfo{author}{\bibfnamefont{R.}~\bibnamefont{Prozorov}},
  \bibnamefont{et~al.}, \bibinfo{journal}{Phys. Rev. B}
  \textbf{\bibinfo{volume}{87}}, \bibinfo{pages}{075120}
  (\bibinfo{year}{2013}).

\end{thebibliography}

\end{document}